\documentclass{arxiv}
\PassOptionsToPackage{dvipsnames}{xcolor}
\usepackage{xcolor}
\usepackage{graphicx,amssymb,amsmath}
\usepackage[colorlinks=true, linkcolor=black, citecolor=black, urlcolor=black]{hyperref}
\usepackage{subcaption}
\usepackage[justification=centering]{caption}
\usepackage{algorithm}
\usepackage{algorithmicx}
\usepackage{algpseudocode}
\usepackage{tikz}
\usepackage{scalerel}
\usepackage{pict2e}
\usepackage{tkz-euclide}
\usetikzlibrary{calc}
\usetikzlibrary{patterns,arrows.meta}
\usetikzlibrary{decorations.markings}
\usetikzlibrary{shadows}
\usetikzlibrary{external}
\usepackage{mathtools}
\usepackage{amsfonts}
\usepackage{pgfplots}
\pgfplotsset{compat=newest}
\usepgfplotslibrary{statistics}
\usepgfplotslibrary{fillbetween}
\usepackage{nicefrac}
\usepackage{subcaption}
\usepackage{pgfkeys}
\usepackage{pgfmath}
\usepackage{tikz-3dplot}
\usepgfplotslibrary{colormaps, patchplots}
\definecolor{DeepSkyBlue}{RGB}{0, 191, 255}
\definecolor{SteelBlue}{RGB}{70, 130, 180}
\definecolor{MidnightBlue}{RGB}{25, 25, 112}
\definecolor{DarkSlateBlue}{RGB}{72, 61, 139}
\definecolor{Blue}{RGB}{0,0,255}
\definecolor{SkyBlue}{RGB}{135,206,235}
\definecolor{NavyBlue}{RGB}{0,0,128}

\title{Improved Wake-Up Time For Euclidean Freeze-Tag Problem}

\author{Sharareh Alipour\thanks{TeIAS, Khatam University, \texttt{sharareh.alipour@gmail.com}}
	\and
	Arash Ahadi\thanks{TeIAS, Khatam University, \texttt{aarash.ahadi.academic@gmail.com}}
	\and
	Kajal Baghestani \thanks{Sharif University of Technology, \texttt{kajal.baghestani@gmail.com}}}
\index{Author, First}
\index{Author, Second}
\index{Author, Third}

\begin{document}
	\thispagestyle{empty}
	\maketitle
	\thispagestyle{empty}

	\begin{abstract}
The Freeze-Tag Problem (FTP) involves activating a set of initially asleep robots as quickly as possible, starting from a single awake robot. Once activated, a robot can assist in waking up other robots. Each active robot moves at unit speed. The objective is to minimize the makespan, i.e., the time required to activate the last robot. A key performance measure is the wake-up ratio, defined as the maximum time needed to activate any number of robots in any primary positions. This work focuses on the geometric (Euclidean) version of FTP in $\mathbb{R}^d$ under the $\ell_p$ norm, where the initial distance between each asleep robot and the single active robot is at most 1. For $(\mathbb{R}^2, \ell_2)$, we improve the previous upper bound of 4.62 (\cite{bonichon:hal-04803161}, CCCG 2024) to 4.31. Note that it is known that 3.82 is a lower bound for the wake-up ratio. In $\mathbb{R}^3$, we propose a new strategy that achieves a wake-up ratio of 12 for $(\mathbb{R}^3, \ell_1)$ and 12.76 for $(\mathbb{R}^3, \ell_2)$, improving upon the previous bounds of 13 and $13\sqrt{3}$, respectively, reported in \cite{alipour}.
\end{abstract}

	\section{Introduction}
	
	The Freeze-Tag Problem (FTP), introduced by Arkin et al.~\cite{ArkinBFMS02}, involves activating a set of $n$ inactive robots as quickly as possible. The process starts with a single active robot, while others remain stationary until they are woken up. Once activated, a robot can assist in waking up other robots. The objective is to minimize the makespan, which is the time required to activate the last robot.
	
	FTP has practical applications in robotics, particularly in swarm control tasks such as environment exploration
	\cite{bruckstein1997probabilistic, gage2001minimum, wagner1999distributed, wagner1998efficiently}, 
	robot formation~\cite{sugihara1996distributed, suzuki1999distributed}, and searching~\cite{wagner1998efficiently}. It is also relevant in network design, including broadcast and IP multicast problems~\cite{arkin2003improved, arkin2006freeze, konemann2005approximating}.
	
	This paper examines the Euclidean (geometric) version of FTP. In this version the input consists of the positions of $n$ asleep robots and one active robot in $\mathbb{R}^d$, along with a distance norm $\ell_p$ for some dimension $d$ and norm parameter $p$. Each robot's position is represented as a point and each active robot moves at unit speed. The objective is to activate all robots. The makespan is defined as the minimum time required to active the last robot; while the wake-up ratio is the maximum makespan among any finite number of robots in any primary positions.
	
   Note that for any point set where the awake robot is within distance \( r \) of all sleeping robots and an upper bound \( c \) exists for the makespan, scaling the unit ball allows constructing a wake-up tree with makespan at most \( r \times c \). Consequently, this results in a \( c \)-approximation algorithm for FTP, as \( r \) is a trivial lower bound on the makespan.

\begin{figure}
	\centering

\begin{tikzpicture}[scale = 0.8]
  \begin{axis}[
    view={290}{30},
    axis lines=middle,
    xlabel={},xlabel={$x$}, ylabel={$y$}, zlabel={$z$},
    xtick={0,1,2}, ytick={0,1,2}, ztick={-2,-1,0},
    grid=major,
    enlargelimits=true,
    scale=0.9,
    z post scale=1.4,
    colormap name=viridis, 
  ]
    
    \addplot3 [patch, faceted color=black, opacity=0.5] coordinates {(1,0,0) (0,1,0) (0,0,1)};
    
    \addplot3 [patch, faceted color=SkyBlue, opacity=0.5] coordinates {(-1,0,0) (0,1,0) (0,0,1)};
    \addplot3 [patch, faceted color=MidnightBlue, opacity=0.85] coordinates {(-1,0,0) (0,-1,0) (0,0,1)};
    
    \addplot3 [patch, faceted color=Blue, opacity=0.5] coordinates {(1,0,0) (0,-1,0) (0,0,-1)};
    
    \addplot3 [patch, faceted color=NavyBlue, opacity=0.5] coordinates {(1,0,0) (0,1,0) (0,0,-1)};
    \addplot3 [patch, faceted color=NavyBlue, opacity=0.5] coordinates {(-1,0,0) (0,1,0) (0,0,-1)};
    \addplot3 [patch, faceted color=DarkSlateBlue, opacity=0.5] coordinates {(-1,0,0) (0,-1,0) (0,0,-1)};
    
  \end{axis}
\end{tikzpicture}

	\caption{The unit ball in $(\mathbb{R}^3, \ell_1)$  }
	\label{ball31}
\end{figure}
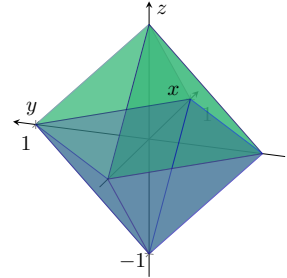

In \( \mathbb{R}^2 \), the unit \( \ell_2 \)-ball is a cyclic disc, while the unit \( \ell_1 \)-ball is a square. In \( \mathbb{R}^3 \), the unit \( \ell_2 \)-ball is a sphere, and the unit \( \ell_1 \)-ball appears as shown in Figure~\ref{ball31}.

	\subsection{Related Work}
    Studies~\cite{abel2017freeze, johnson2017easier, pedrosa2023freeze} confirm that FTP is NP-hard in $(\mathbb{R}^3, \ell_p)$ for all $p \geq 1$. Similarly, FTP is NP-hard for $(\mathbb{R}^2, \ell_2)$, though complexity results for other norms remain unresolved~\cite{abel2017freeze}. It is suspected that FTP is also NP-hard for $(\mathbb{R}^2, \ell_1)$~\cite{arkin2006freeze}.
	
	Let $\gamma_{d,p}$ be the wake-up time in $(\mathbb{R}^d, \ell_p)$.
	Despite the complexity of the problem, few studies focus on the wake-up ratio. Bonichon et al. showed that for 
	$(\mathbb{R}^2, \ell_1)$, it is at most $5$, leading to an upper bound of 
	$5\sqrt{2}$ for $(\mathbb{R}^2,\ell_2)$ \cite{BonichonCGH24}. 
	Consider an instance for FTP in $(\mathbb{R}^2, \ell_2)$ containing exactly 4 robots located at $(0,1)$, $(0,-1)$, $(1,0)$, and $(-1,0)$. It is easy to see that the makespan of this instance is $1+2\sqrt{2}\approx 3.83$. In \cite{BonichonCGH24} it is conjectured that this input takes the maximum makespan among all instances; in other words $\gamma_{2, 2}\approx 3.83$.  
	Recently, the upper bound for 
	$\gamma_{2, 2}$ was reduced to $5.41$ by \cite{alipour} and $4.62$ by \cite{bonichon:hal-04803161}. 

    Arkin et al.~\cite{ArkinBFMS02} formulated the FTP as a problem of constructing a rooted spanning tree with minimized weighted depth. The root represents the initially active robot, which has one child, while the \( n \) inactive robots serve as nodes with at most two children each. Edges represent distances in the metric space. This structure, known as a wake-up tree, defines the wake-up time as its weighted depth. In Subsection \ref{aaa} we use this spanning tree.

    For $\gamma_{3, 1}$ Alipour et al. proposed an algorithm with a wake-up ratio of $13$, which corresponds to $13\sqrt{3}$ for $\gamma_{3, 2}$ \cite{alipour}. See table \ref{table}.

	\subsection{Our Contribution}

    For the Freeze-Tag problem in \( (\mathbb{R}^2, \ell_2) \), we use the crown strategy from \cite{alipour, bonichon:hal-04803161} with a more detailed and slightly more complex version. For further details, we propose algorithms based on the positions of the three nearest asleep robots to the center; for each FTP instance, we select the one that minimizes the makespan.

In Section \ref{ix}, we present the upper bound for \( \gamma_{2, 2} \); and in Section \ref{section3d}, we present the upper bound of 12 for \( \gamma_{3, 1} \). The upper bound of 12.76 for \( \gamma_{3, 2} \) is proved in Appendix $\mathbf{C}$. See Table \ref{table} for a summary of the results.

	\begin{table}[h]
		\centering
		\begin{tabular}{|c|c|c|}
			\hline
			(Dim, Norm)  & Previous Bound  & New Bound \\
			\hline 
			$(\mathbb{R}^2, \ell_2)$ & 4.62 \,\,\,\,\,\,\cite{bonichon:hal-04803161} & 4.31 \,\,\,\,(Thm. \ref{4.3})\\
			$(\mathbb{R}^3, \ell_1)$ & 13 \,\,\,\,\,\,\,\,\, \cite{alipour} & 12\,\,\,\,\,\,\,\,\,\,\,(Thm. \ref{norm1}) \\
			$(\mathbb{R}^3, \ell_2)$ & 22.52 \,\, \cite{alipour}& 12.76 \,\,\,(Thm. \ref{norm2})\\
			\hline
		\end{tabular}
		\caption{Summary of results.}
		\label{table}
	\end{table}

	\section{FTP in  $(\mathbb{R}^2, \ell_2)$} \label{ix}
In this section we show that \( 4.31 \) is an upper bound for the wake-up ratio of the FTP in \( (\mathbb{R}^2, \ell_2) \).

\begin{theorem} \label{4.3}
		$\gamma_{2,2} \leq 4.31$.
\end{theorem}

    In Subsection \ref{aaa}, we present several algorithms for solving the Freeze-Tag Problem. For any given FTP instance, we select the algorithm that yields the smallest makespan among the proposed options. We then use a computer program to compute the makespan across a range of input instances. Although the space of possible inputs for FTP is infinite, our program evaluates only a finite subset. However, we carefully select representative cases such that every arbitrary input has a makespan close to that of one of the tested instances. As a result, the program introduces a bounded approximation error. In Appendix $\mathbf{B}$, we provide a detailed analysis of this approximation error.

    We begin by introducing some concepts and notations. Consider a Freeze-Tag instance with $n$ asleep robots, denoted $p_1, \dots, p_n$, and a single active robot $p_0$ located at the origin in $(\mathbb{R}^2, \ell_2)$. Let $r_i = |p_i - O|_2$ denote the Euclidean distance from $p_i$ to the origin $O$, with the assumption that $r_1 \leq r_2 \leq \dots \leq r_n$. A circle $C(r)$ refers to the circle centered at $O$ with radius $r$.

    We define a crown, denoted by $\mathcal{R}(1 - r, \theta)$, as a sector-shaped region bounded by an inner radius $r$, an outer radius 1, and a central angle $\theta$ (see Figure~\ref{crown}). It consists of all points $p_i$ such that $r \leq |p_i - O|_2 \leq 1$ and whose angular coordinates fall within the sector defined by $\theta$. The width of the crown, denoted by $w$, is given by $w = 1 - r$.

    We use $|a - b|$ to denote the absolute value of the difference between $a$ and $b$. Throughout this paper, we use $p_i$ to refer both to the robot and its initial position.

	\subsection{Algorithms} \label{aaa}
	We present several algorithms, each of which outperforms the others in specific instances. All of our algorithms are based on the crown strategy. 
	In all algorithms we use the following lemmas from \cite{bonichon:hal-04803161}. 
	Let $\phi$ denote the golden ratio, defined as $\frac{1+\sqrt{5}}{2}$.
	
	\begin{lemma} (Corollary 1 of \cite{bonichon:hal-04803161})\label{phi}
		There exists a strategy to wake up all of the robots in a crown of angle $\Theta$ and width $w$ starting
		with one awake robot at a corner in time at most $\Theta+\phi w$.
	\end{lemma}
	
	\begin{lemma} \label{up} (Lemma 5 of \cite{bonichon:hal-04803161}) There exists a strategy to wake up all of the
		robots in a crown of angle $\Theta$ and width $w$ starting with
		two awake robots at a corner on the exterior side of the
		crown in time at most $\Theta + \Big(\frac{\phi ^4}{\phi^3 + \Theta}\Big)w$.
	\end{lemma}
	
  Lemma \ref{crown} is an extension of Corollary 3 from \cite{bonichon:hal-04803161}. Their proof relies on Proposition 14 from \cite{BonichonCGH24} and applies only to angles less than or equal to \( \pi \). In Lemma \ref{crown}, we demonstrate that the statement holds for any arbitrary angle. The proof is in Appendix $\mathbf{A}$.

	\begin{lemma}
		\label{crown}
		There exists a strategy to wake up all of the robots in a crown of angle $\Theta$ and width $w$
		starting with one awake robot at a corner on the interior side of the crown in time at most $\Theta + \Big(1+\frac{\phi ^4}{\phi^3 + \Theta} \Big)w$.
	\end{lemma}

	\begin{figure}
        \centering
\begin{tikzpicture}[scale = 0.7]

\pgfmathsetmacro{\Theta}{38.8}  
\pgfmathsetmacro{\r}{4}  

    \coordinate (A) at (5,4);
    \coordinate (B) at (4.5,3.2);
    \coordinate (C) at (4,2.4);
    \coordinate (a1) at (1.05,2.45);
    \coordinate (a2) at (1.15,2.5);
    \pgfmathsetmacro{\a}{-2.7}
    \pgfmathsetmacro{\b}{6.75}
    \pgfmathsetmacro{\am}{-1.653}
    \pgfmathsetmacro{\bm}{4.1352}
    \pgfmathsetmacro{\aa}{-4.76}
    \pgfmathsetmacro{\bb}{11.9}


    \draw (A) arc(90-\Theta:90+\Theta:\r);


    \draw (C) arc(90-\Theta:90+\Theta: 0.6 * \r);

    \draw [thick, color={rgb,255:red,255;green,80;blue,0}] (B) -- (5,4)
    node[pos=0.5, below right , scale=1] { $\alpha(1-r)$};

    \draw [thick, color={rgb,255:red,255;green,165;blue,0}] (C) -- (B)
    node[pos=0.5, below right , scale=1] {$(1-\alpha)(1-r)$};
    \draw[dash pattern=on 2pt off 2pt, thick, color={rgb,255:red,0;green,100;blue,200}] (2.5,0) -- (C)
     node[pos=0.5, below right , scale=1] {$r$};

    \draw  (0.5,3.2) -- (0,4);
    \draw (1,2.4) -- (0.5,3.2);
    \draw[dash pattern=on 2pt off 2pt] (2.5,0) -- (1,2.4);

    \draw[dash pattern=on 2pt off 2pt] (2.5 , \aa * 2.5 + \bb) -- (1.9 , \aa * 1.9 +\bb );


    \draw[
          color={rgb,255:red,200;green,10;blue,100},thick] (1.9 , \aa * 1.9 + \bb)--(1.56 , \aa * 1.56 + \bb) ;

    \draw[color={rgb,255:red,200;green,10;blue,100},thick] (a1) arc(126:106.3:2.7);

    \draw[thick,->,color={rgb,255:red,200;green,10;blue,100}]  (1.56 , \aa * 1.56 + \bb) -- (1.5 , \aa * 1.5 +\bb );


    \draw (2.63,0.2) arc (90-\Theta:90+\Theta:\r * 0.05) node[yshift = 5, xshift = 4,scale = 1] {  
 $\theta$};

    \draw (2.1 , \aa * 2.1 + \bb) arc (95:90+\Theta:\r * 0.28) node[yshift = 8, xshift = 2,scale = 1] { 
 $\gamma$};

    \node at (2.5,0) [yshift = -5, scale=1.3] {$O$}; 

    \node at (a1) [circle, fill, inner sep=1pt] {};  
    \node at (a1) [left, scale=1.1] {$a$}; 

    \node at (1.6 , \aa * 1.6 + \bb) [circle, fill, inner sep=1pt] {};
    \node at (1.6 , \aa * 1.6 + \bb) [xshift=10pt,
    yshift = -1pt ,scale=1.1] {$b$}; 

\end{tikzpicture}
        \caption{Activating a crown by one activated robot on the inner corner}
        \label{x}
    \end{figure}
	
    First, using a small improvement of algorithm of \cite{bonichon:hal-04803161}, we improve the bound of 4.62 to 4.54. Next by some other methods we decrease this value to 4.31. 
	\subsection*{A Small Improvement of The Algorithm of \cite{bonichon:hal-04803161}}

	In \cite{bonichon:hal-04803161}, the bound $4.62$ has been obtained by trade-off between two strategies. 
	In the STRATEGY 1, \( p_0 \) first moves toward \( p_1 \). Once \( p_1 \) is activated, both $p_0$ and $p_1$ move to the origin $O$. Next, \( p_0 \) moves toward \( p_2 \) and activates it, while $p_1$ moves to a point $t$ in the unit circle such that $\angle{p_2Ot}=\frac{2}{3}\pi$ and $\|O-t\|=r_2$. Finally each of $p_0, p_1$ and $p_2$ activates a crown $\mathcal{R}(1-r_2, \frac{2}{3}\pi)$.
	
	Now we improve this strategy. Our improved strategy is denoted by three-crowns.
	Note that the step of coming back to the origin $O$ can be improved. In fact, $p_0$ can move directly to $p_2$ and also $p_1$ can move directly to $t$. Next each of $p_0, p_1$ and $p_2$ activates a crown.
	However we should compute the optimal angle for the crowns that should be activated by these three points. 
	
	Let the angles of each of  $\mathcal{R}_0$ and $\mathcal{R}_2$ be $x$; so the angle of $\mathcal{R}_1$ is $2\pi - 2x$. By Lemma \ref{crown}, the total time of activating $\mathcal{R}_0$ and $\mathcal{R}_2$ is 
	$$T_{0,2} = r_1+||p_1-p_2||_2+ x + \Big(1+\frac{\phi ^4}{\phi^3 + x} \Big)(1-r_2).$$
	Also, by Lemma \ref{crown} the total time of activating $\mathcal{R}_1$ is 
	$$T_1=r_1+||p_1-t||_2+ (2\pi-2x) + \Big(1+\frac{\phi ^4}{\phi^3 + 2\pi-2x} \Big)(1-r_2).$$
	
	We set $x$ such that the maximum of $T_{0, 2}$ and $T_1$ takes the minimum possible value (in this case, $T_{0, 2}=T_1$). Thus, the three-crowns algorithm activates all robots in at most 
	$$\tau_1:=\min\limits_{0\leq x \leq \pi} \max \{T_{0, 2}, T_1\}.$$

	In the STRATEGY 2 of \cite{bonichon:hal-04803161}, first $p_0$ moves to $p_1$; next $p_0$ starts to activate a crown $\mathcal{R}_0=\mathcal{R}(1-r_1, \pi)$ and $p_1$ starts to activate a crown $\mathcal{R}_1=\mathcal{R}(1-r_1, \pi)$; such that the union of $\mathcal{R}_0, \mathcal{R}_1$ and $C(r_1)$ is the unit cycle. Since there is no asleep robot in $C(r_1)$, this strategy activates all robots. By Lemma \ref{crown}, the total time of this strategy is
	$$\tau_2:=r_1 + \pi + \Big(1+\frac{\phi ^4}{\phi^3 + \pi} \Big)(1-r_1).$$
	We call this strategy two-crowns algorithm.
	
	The wake-up ratio of 4.62 is found by solving a max-min optimization. For each FTP input, the make-span is calculated using two strategies, and the one with the smallest make-span is chosen. The input with the largest make-span has a value of 4.62. However, by using the three-crowns algorithm (a modified version of STRATEGY 1) and the two-crowns algorithm (which is the same as STRATEGY 1), we can achieve a wake-up ratio of $\min \{\tau_1, \tau_2\} = 4.54$, where the minimum is given over all values of $0 \leq r_1 \leq r_2 \leq 1$.
	
	\subsection*{\textbf{Two-crowns algorithm with respect to $r_3$}}
	In the two-crowns algorithm, the active robot \( p_0 \) at the origin first moves toward \( p_1 \). Once \( p_1 \) is activated, there are now two active robots at its position. These robots then move \( r_3 - r_1 \) units in the direction of \( \overrightarrow{Op_1} \).
	Next, the crown \( \mathcal{R}(1 - r_3, 2\pi) \) is divided into two crowns
	\[
	\mathcal{R}_{\text{left}} = \mathcal{R}(1 - r_3, 2\pi - x) \, \quad \text{and} \,\quad  \mathcal{R}_{\text{right}} = \mathcal{R}(1 - r_3, x)
	\]
	
	both starting from the radius that contains \( p_1 \). The value of \( x \) is determined later. One of these crowns is activated by \( p_0 \), and the other by \( p_1 \). Since there are no robots, except \( p_2 \), inside the circle \( C(r_3) \), by activating \( p_2 \) and both \( \mathcal{R}_{\text{left}} \) and \( \mathcal{R}_{\text{right}} \), the entire unit circle is activated.
	
	\begin{figure}
    \centering
\begin{tikzpicture}[scale = 1.25]
    \coordinate (O) at (2,2);
    \coordinate (A) at (0.5,2);
    \coordinate (B) at (3.33,3.5);
    \coordinate (C) at (2,0);
    \coordinate (a) at (2.14,2.28);
    \pgfmathsetmacro{\a}{2}
    \pgfmathsetmacro{\b}{-2}
    \pgfmathsetmacro{\am}{1}
    \pgfmathsetmacro{\bm}{0}
    \pgfmathsetmacro{\r}{0.7}
    \coordinate (p1) at (1.65,1.65);
    \coordinate (p2) at (1.25,2.75);
    \coordinate (p3p1) at (0.95,\am*0.95+\bm);
    \coordinate (p3p2) at (0.95,-1*0.95+4);


    \draw (2,2) circle (2);
    \draw [dash pattern=on 2pt off 2pt] (2,2) circle (1.5);


    \draw[color={rgb,255:red,0;green,100;blue,200}] (O) -- (2.67,\a*2.67+\b)
    node[pos=0.5, left , scale=1] {$r_3$};
    
    \draw[color={rgb,255:red,135;green,206;blue,235}] (2.67,\a*2.67+\b) -- (2.9,\a*2.9+\b)
    node[pos=0.5, yshift = 3,xshift = -17 , scale=1] {$1-r_3$};

    \draw[postaction={decorate},
          decoration={markings, mark=at position 0.9 with {\arrow{>}}},
           color={rgb,255:red,56;green,73;blue,24},thick] (O) -- (p1);

    \draw[postaction={decorate},
          decoration={markings, mark=at position 1 with {\arrow{>}}},
           color={rgb,255:red,56;green,73;blue,24},thick] (p1) -- (p3p1);

    \draw[dash pattern=on 2pt off 2pt,  color = gray] (p3p1) -- (0.59,\am*0.59+\bm);


    \draw[postaction={decorate},
          decoration={markings, mark=at position 0.9 with {\arrow{>}}},
           color={rgb,255:red,56;green,73;blue,24},thick] (1.3,2.8) -- (1,3.1);
    \draw[postaction={decorate},
          decoration={markings, mark=at position 0.82 with {\arrow{>}}},
           color={rgb,255:red,56;green,73;blue,24},thick] (p3p2) -- (p2);


    \node at (O) [circle, fill, inner sep=1pt] {};
    \node at (O) [xshift=5pt,
    yshift = -5pt ,scale=1] {$p_0$}; 

    \node at (p1) [circle, fill, inner sep=1pt] {};
    \node at (p1) [xshift= 5pt,
    yshift = -5pt ,scale=1] {$p_1$}; 

    \node at (p2) [circle, fill, inner sep=1pt] {};
    \node at (p2) [xshift=7pt,
    yshift = -2pt ,scale= 1] {$p_2$}; 

    \node at (p3p2) [circle, fill, inner sep=1pt] {};
    \node at (p3p2) [xshift=-8pt,
    yshift = 2pt ,scale= 1] {$p^*_2$}; 

    \draw (a) arc (70:222:0.31) node[left, above,xshift=-10pt, yshift=6pt] {$x$};

    \draw[postaction={decorate},
          decoration={markings, mark=at position 1 with {\arrow{>}}},
          color={rgb,255:red,200;green,10;blue,100},thick] (p3p1) arc(180:225:\r);

    \draw[postaction={decorate},
          decoration={markings, mark=at position 1 with {\arrow{>}}},
          color={rgb,255:red,200;green,10;blue,100},thick] (p3p1) arc(270:225:\r);

    \end{tikzpicture}

    \caption{Two-crowns algorithm with respect to $r_3$}
    \label{2crowns3}
\end{figure}
	
	Without loss of generality, suppose that robot \( p_2 \) is on the right side of \( \overrightarrow{Op_1} \). See Figure \ref{2crowns3}. Consider an instance \( p_1, p_2^*, p_3, \dots, p_n \) of the FTP, similar to \( p_1, p_2, \dots, p_n \), with one difference: the two robots \( p_2 \) and \( p_2^* \) and the point $O$ belong to a line, and also \( \|O-p_2^*\|_2 = r_3 \). By Lemma \ref{crown}, the crown \( \mathcal{R}_{\text{right}} \) can be activated in 
	
	\[
	x + \left( 1 + \frac{\phi^4}{\phi^3 + x} \right) (1-r_3)
	\]
	
	time units. 
	We now present a strategy to activate \( p_2 \) and all robots except \( p_2^* \) in \( \mathcal{R}_{\text{right}} \). Consider the wake-up tree of activating the crown \( \mathcal{R}_{\text{right}} \) by the strategy of Lemma \ref{crown}; the parent of \( p_2^* \) activates \( p_2 \). For more details, the parent of \( p_2^* \) first moves to the position of \( p_2^* \), then moves to \( p_2 \), and finally both  \( p_2 \) and the parent move to the position of \( p_2^* \). Other nodes and edges of the wake-up tree do not change.
	
	Since \( \|p_2^*- p_2\|_2 = r_3 - r_2 \), this strategy takes
	
	\[
	T_{\text{right}} = x + \left( 1 + \frac{\phi^4}{\phi^3 + x} \right) (1-r_3) + 2(r_3 - r_2)
	\]
	
	time units.
	
	Activating \( \mathcal{R}_{\text{left}} \) by \( p_1 \) spends 
	
	\[
	T_{\text{left}} = (2\pi-x)+ \left( 1 + \frac{\phi^4}{\phi^3 + 2\pi - x} \right) (1-r_3)
	\]
	
	time units. Also moving \( p_0 \) and \( p_1 \) to a corner of their crowns spends \( r_3 \) time units.

	We set the angle \( x \) such that the maximum of \( T_{\text{right}} \) and \( T_{\text{left}} \) is minimized (in this case, \( T_{\text{right}} = T_{\text{left}} \)). Consequently, the total time of the algorithm is
	
	\[
	r_3 + \min_
	{0\leq x\leq \pi}\max \{ T_{\text{right}}, T_{\text{left}} \}.
	\]
	
	\subsection*{\textbf {Three-crowns algorithm with respect to $r_3$}}
	
	In the three-crowns algorithm, \( p_0 \) first moves toward \( p_1 \). Once \( p_1 \) is activated, there are two active robots at its position. Next, \( p_1 \) moves toward \( p_2 \) and activates it. Then, both \( p_1 \) and \( p_2 \) move outward along \( \overrightarrow{Op_2} \) for \( r_3 - r_2 \) units. 
	
	At this point, each of \( p_1 \) and \( p_2 \) activates a crown: \( p_1 \) activates a crown \( \mathcal{R}_1 = \mathcal{R}(1-r_3, x) \) and \( p_2 \) activates a crown \( \mathcal{R}_2 = \mathcal{R}(1-r_3, x) \), where the value of \( x \) will be determined later.
	
	Meanwhile, \( p_0 \) continues to activate robots in the remaining crown \( \mathcal{R}(1 - r_3,  2\pi - 2x) \) as follows. Let \( q \) be the point such that \( \angle p_2 O q = x \) and \( \|O - q\|_2 = r_3 \). In this case, \( p_0 \) moves toward \( q \); next, \( p_0 \) activates a crown \( \mathcal{R}_0 = \mathcal{R}(1 - r_3, \pi - 2x) \). See Figure \ref{3crowns3}.

    \begin{figure}
        \centering

    \begin{tikzpicture}[scale = 1.3]
    \coordinate (O) at (2,2);
    \coordinate (A) at (0.5,2);
    \coordinate (B) at (3.33,3.5);
    \coordinate (C) at (2,0);
    \coordinate (a) at (2.14,2.28);
    \pgfmathsetmacro{\a}{0.25}
    \pgfmathsetmacro{\b}{1.5}
    \pgfmathsetmacro{\am}{0.25}
    \pgfmathsetmacro{\bm}{1.5}
    \pgfmathsetmacro{\aa}{-0.25}
    \pgfmathsetmacro{\bb}{2.5}
    \pgfmathsetmacro{\r}{0.7}
    \pgfmathsetmacro{\rp}{0.5}
    \coordinate (p1) at (1.75,2.52);
    \coordinate (p2) at (2,3);
    \coordinate (p3) at (3.12,\a*3.12+\b);
    \coordinate (pi) at (2.7,\aa*2.7+\bb);
    \coordinate (p2aim1) at (2,3.25);
    \coordinate (q) at (0.78,\a * 0.78+\b);


    \draw (2,2) circle (2);
    \draw [dash pattern=on 2pt off 2pt] (2,2) circle (1.25);

    \draw (2.3,\aa*2.3+\bb) arc(-5.5:77:0.4)
    node[pos=0.5, right , scale=1] {\small $x$};

    \draw (2,2.2) arc(90:190:0.2)
    node[pos=0.5, left , scale=1] {\small $x$};

    
    \draw[dash pattern=on 2pt off 2pt] (O) -- (0.78,\a * 0.78+\b);
    \draw[dash pattern=on 2pt off 2pt] (0.78,\am* 0.78+\bm) -- (0.07,\a*0.07+\b);


    \draw[dash pattern=on 2pt off 2pt] (O) -- (2,3.25);

    \draw[dash pattern=on 2pt off 2pt] (2,3.25) -- (2,4);


    \draw[color={rgb,255:red,0;green,100;blue,200}] (O) -- (3.226,\aa*3.226+\bb)
    node[pos=0.5, below , scale=1] {\small $r_3$};
    
    \draw[color={rgb,255:red,135;green,206;blue,235}] (3.226,\aa*3.226+\bb) -- (3.93,\aa*3.93+\bb)
    node[pos=0.5, below , scale=1,xshift = -2pt] {\small $1-r_3$};


    \draw[postaction={decorate},
          decoration={markings, mark=at position 0.7 with {\arrow{>}}},
           color={rgb,255:red,56;green,73;blue,24},thick] (O) -- (p1);
    \draw[postaction={decorate},
          decoration={markings, mark=at position 0.82 with {\arrow{>}}},
           color={rgb,255:red,56;green,73;blue,24},thick] (p1) -- (p2);
    \draw[postaction={decorate},
          decoration={markings, mark=at position 0.82 with {\arrow{>}}},
           color={rgb,255:red,56;green,73;blue,24},thick] (p2) -- (p2aim1);
    \draw[postaction={decorate},
          decoration={markings, mark=at position 0.82 with {\arrow{>}}},
           color={rgb,255:red,56;green,73;blue,24},thick] (p1) -- (q);

    \draw[postaction={decorate},
          decoration={markings, mark=at position 1 with {\arrow{>}}},
          color={rgb,255:red,200;green,10;blue,100},thick] (p2aim1) arc(135:90:\r);

    \draw[postaction={decorate},
          decoration={markings, mark=at position 1 with {\arrow{>}}},
          color={rgb,255:red,200;green,10;blue,100},thick] (p2aim1) arc(45:90:\r);

    \draw[postaction={decorate},
          decoration={markings, mark=at position 1 with {\arrow{>}}},
          color={rgb,255:red,200;green,10;blue,100},thick] (q) arc(145:190:\r);

    \node at (O) [circle, fill, inner sep=1pt] {};
    \node at (O) [xshift=0pt,
    yshift = -15pt ,scale=1] {\small $p_0$}; 

    \node at (p1) [circle, fill, inner sep=1pt] {};
    \node at (p1) [xshift= -9pt,
    yshift = 3pt ,scale=1] {\small $p_1$}; 

    \node at (p2) [circle, fill, inner sep=1pt] {};
    \node at (p2) [xshift=8pt,
    yshift = -2pt ,scale=1] {\small $p_2$}; 
    
    \node at (q) [circle, fill, inner sep=1pt] {};
    \node at (q) [xshift=8pt,
    yshift = -4pt ,scale=1] {\small $q$}; 

    \end{tikzpicture}

        \caption{Three-crowns algorithm with respect to $r_3$}
        \label{3crowns3}
    \end{figure}

	By Lemma \ref{crown}, the total time of activating \( \mathcal{R}_1 \) and \( \mathcal{R}_2 \) is
	\vspace{-0.2 cm}
	  \begin{align}
	T_{1,2} & = r_1 + \|p_1- p_2\|_2 + (r_3 - r_2)  \notag\\
	& + \left(x+\left(1+\frac{\phi^4}{\phi^3 + x}\right)(1-r_3) \right). \notag
	\end{align}
	
	Also the total time of activating $\mathcal{R}_0$ is
	
    \begin{align}
    T_0 &= r_1 + \|p_1 - q\|_2 + \Big( (2\pi - 2x) \notag \\
    &\quad + \Big(1 + \frac{\phi^4}{\phi^3 + 2\pi - 2x}\Big)(1 - r_3) \Big). \notag
    \end{align}

	We set the value \( x \) such that \( T_{1,2} = T_0 \). Consequently, the total time of the three-crowns algorithm is
	
	\[\min\limits_{0 \leq x \leq \pi} \max \{T_{1,2}, T_0\}.\]
	
	\subsection*{\textbf{(Two or Four)-Crowns Algorithm}}

	First, we outline the overall structure of the four-crowns algorithm. In this algorithm, \( p_0 \) activates a robot \( q \). Then, both \( p_0 \) and \( q \) activate new robots, denoted as \( p' \) and \( q' \), respectively. Finally, the ring is divided into the crowns \( \mathcal{R}_{p_0} \), \( \mathcal{R}_{p'} \), \( \mathcal{R}_{q} \), and \( \mathcal{R}_{q'} \), each assigned to one of \( p_0 \), \( p' \), \( q \), and \( q' \). Since \( p_0 \) and \( p' \) start their crowns simultaneously, the angles of \( \mathcal{R}_{p_0} \) and \( \mathcal{R}_{p'} \) should be equal in the optimum case. However, if \( \angle p' O q' = \pi - \beta \) for some \( \beta > 0 \), then before starting to activate the crowns, some of those 4 robots should move to new positions \( p'' \) and \( q'' \) such that \( \angle p'' O q'' = \pi \). After this adjustment, the robots can start activating the crowns.

	Let \( p^\dagger \) be the robot that is farthest in angular distance from \( p_1 \) with the angle \( \angle p_1Op^\dagger = \pi - \beta \) for some \( \beta \). As discussed earlier, if \( \beta \) is small, the four-crowns algorithm may achieve an efficient wake-up time; however, this is not the case when \( \beta \) is large. In such case, the two-crowns algorithm  is more sufficient. Specifically, it is enough to remove the region with angle \( \beta \) from both crowns, as no robots are located in these areas. In other words, in the two-crowns algorithm, two crowns with an angle of \( \pi - \beta \) should be activated instead of the two crowns with an angle of \( \pi \). So, based on the value of $\beta$, we decide to choose one of the algorithms (two or four)-crowns.
    These algorithms are discussed in detail in the next two parts.
	
	\textbf{Four-Crowns Algorithm with Respect to $\beta$} 
	
	In this algorithm, first \( p_0 \) moves to \( p_2 \). Next, \( p_0 \) moves to \( p_1 \), and \( p_2 \) moves to \( p^\dagger \), where \( p^\dagger \) is the robot that is farthest in angular distance from \( p_1 \); let $\beta$ be an angle such that \( \angle p_1Op^\dagger = \pi - \beta \).
	
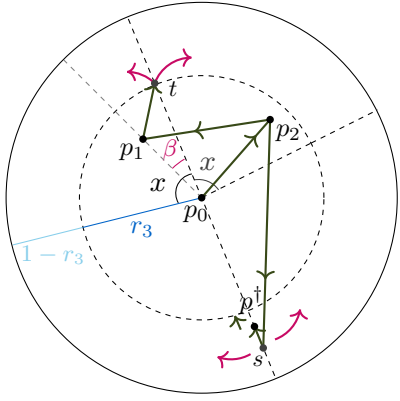
\begin{figure}
        \centering

  \begin{tikzpicture}[scale = 1.3]
    \coordinate (O) at (2,2);
    \coordinate (A) at (0.5,2);
    \coordinate (B) at (3.33,3.5);
    \coordinate (C) at (2,0);
    \coordinate (a) at (2.14,2.28);
    \pgfmathsetmacro{\a}{1/2}
    \pgfmathsetmacro{\b}{1}
    \pgfmathsetmacro{\am}{0.25}
    \pgfmathsetmacro{\bm}{1.5}
    \pgfmathsetmacro{\aa}{-2.44}
    \pgfmathsetmacro{\bb}{6.88}
    \pgfmathsetmacro{\aam}{-1}
    \pgfmathsetmacro{\bbm}{4}
    \pgfmathsetmacro{\rThree}{1.25}
    \coordinate (p1) at (1.4,2.6);
    \coordinate (p1aim) at (1.52,\aa*1.52 +\bb);
    \coordinate (p2) at (2.7,2.8);
    \coordinate (p3) at (3.12,\a*3.12+\b);
    \coordinate (pi) at (2.54,\aa*2.54+\bb);
    \coordinate (s) at (2.63,\aa*2.63+\bb);
    \coordinate (t) at (1.52,\aa*1.52+\bb);


    \draw (2,2) circle (2);
    \draw [dash pattern=on 2pt off 2pt] (2,2) circle (\rThree);

    \draw[color = darkgray] (1.925,1.925*\aa+\bb) arc(100:35:0.25)
    node[pos=0.5, above, scale=1.2] {\small $x$};

    \draw (1.9,1.9*\aa+\bb) arc(100:210:0.2)
    node[pos=0.5, left, scale=1.2] {\small $x$};

    \draw[color={rgb,255:red,200;green,10;blue,100}] (1.7,2.3) arc(135:120:0.6)
    node[pos=0.5,xshift = -3pt,yshift = 5pt, scale=1] {\small $\beta$};

    \draw[dash pattern=on 2pt off 2pt, color = gray] (O) -- (0.59,\aam*0.59+\bbm);

    \draw[dash pattern=on 2pt off 2pt] (O) -- (3.12,\a*3.12+\b);

    \draw[dash pattern=on 2pt off 2pt] (3.12,\a*3.12+\b) -- (3.8,\a*3.8+\b);


    \draw[color={rgb,255:red,0;green,100;blue,200}] (O) -- (0.78,\am * 0.78+\bm)
    node[pos=0.5, below , scale=1] {$r_3$};
8 
    \draw[color={rgb,255:red,135;green,206;blue,235}] (0.78,\am* 0.78+\bm) -- (0.07,\am*0.07+\bm)
    node[pos=0.5, yshift = -7, scale=1,xshift = 2pt] {\small $1-r_3$};

    \draw[dash pattern=on 2pt off 2pt]  (1.25,\aa*1.25+\bb) -- (2.762,\aa*2.762+\bb);

    \draw[postaction={decorate},
          decoration={markings, mark=at position 0.8 with {\arrow{>}}},
           color={rgb,255:red,56;green,73;blue,24},thick] (O) -- (p2);
    \draw[postaction={decorate},
          decoration={markings, mark=at position 0.6 with {\arrow{>}}},
           color={rgb,255:red,56;green,73;blue,24},thick] (p2) -- (p1);
           
    \draw[postaction={decorate},
          decoration={markings, mark=at position 0.7 with {\arrow{>}}},
           color={rgb,255:red,56;green,73;blue,24},thick] (p2) -- (s);

    \draw[postaction={decorate},
          decoration={markings, mark=at position 1 with {\arrow{>}}},
           color={rgb,255:red,56;green,73;blue,24},thick] (p1) -- (p1aim);

    \draw[postaction={decorate},
          decoration={markings, mark=at position 1 with {\arrow{>}}},
           color={rgb,255:red,56;green,73;blue,24},thick] (s) -- (pi);

    \draw[postaction={decorate},
          decoration={markings, mark=at position 1 with {\arrow{>}}},
           color={rgb,255:red,56;green,73;blue,24},thick] (2.4,0.65) -- (2.34,0.79);

    \draw[postaction={decorate},
          decoration={markings, mark=at position 1 with {\arrow{>}}},
          color={rgb,255:red,200;green,10;blue,100},thick] (p1aim) arc(160:90:0.4);

    \draw[postaction={decorate},
          decoration={markings, mark=at position 1 with {\arrow{>}}},
          color={rgb,255:red,200;green,10;blue,100},thick] (p1aim) arc(45:90:0.4);

    \draw[postaction={decorate},
          decoration={markings, mark=at position 1 with {\arrow{>}}},
          color={rgb,255:red,200;green,10;blue,100},thick] (2.75,0.55) arc(290:350:0.4);

    \draw[postaction={decorate},
          decoration={markings, mark=at position 1 with {\arrow{>}}},
          color={rgb,255:red,200;green,10;blue,100},thick] (2.5,0.4) arc(300:250:0.4);

    \node at (O) [circle, fill, inner sep=1pt] {};
    \node at (O) [xshift=-2pt,
    yshift = -6pt ,scale=1] {$p_0$}; 

    \node at (p1) [circle, fill, inner sep=1pt] {};
    \node at (p1) [xshift= -4pt,
    yshift = -5pt ,scale=1] {$p_1$}; 

    \node at (p2) [circle, fill, inner sep=1pt] {};
    \node at (p2) [xshift=7pt,
    yshift = -5pt ,scale=1] {$p_2$}; 

    \node at (pi) [circle, fill, inner sep=1pt] {};
    \node at (pi) [xshift=-1pt,
    yshift = 10pt ,scale=1] {$p^\dagger$}; 

    \node at (s) [circle, fill, inner sep=1pt,color = darkgray] {};
    \node at (s) [xshift=-2pt,
    yshift = -5pt ,scale=0.9] {$s$}; 

    \node at (t) [circle, fill, inner sep=1pt,color = darkgray] {};
    \node at (t) [xshift=7pt,
    yshift = -2pt ,scale=0.9] {$t$}; 

    \end{tikzpicture}

        \caption{Four crowns are activated according to the four-crowns algorithm}
        \label{24crowns}
    \end{figure}

	Let \( t \) be a point where \( \angle p^\dagger O t = \pi \) and \( \|t-O\|_2 = r_3 \). In the next step, both \( p_0 \) and \( p_1 \) move toward \( t \) to activate their crowns, \( \mathcal{R}_0 = \mathcal{R}(1 - r_3, x) \) and \( \mathcal{R}_1 = \mathcal{R}(1 - r_3, x) \), respectively. In order to activate the two other crowns  \(\mathcal{R}_2= \mathcal{R}(1 - r_3, \pi - x) \) and  \( \mathcal{R}_\dagger=\mathcal{R}(1 - r_3, \pi - x) \) (those of \( p_2 \) and \( p^\dagger \)), 
	note that if the robot \( p^\dagger \) is positioned near the inner arc of \( \mathcal{R}_\dagger\),
	it is more efficient for \( p_2 \) and \( p^\dagger \) to start their crowns from the inner corner of the crown rather than from the outer corner, i.e., we consider the corner of these two crowns which is the
	nearest one to \(p^\dagger\) and in the worst case \(p^\dagger\) is
	exactly in the middle, so we consider this case and first we go to the middle point \( s \) and then continue to the corner which is closer to \( p^\dagger \) (in this way we activate \( p^\dagger \) too).
	In other words, \( s \) is a point on the radius \( Op^\dagger \), with a distance of \( r_3 + \frac{1 - r_3}{2} \) from \( O \). Next, \( p_2 \) moves to \( p^\dagger \) in the direction of \( \overrightarrow{Op^\dagger} \) or \( \overleftarrow{Op^\dagger} \). Afterward, both \( p_2 \) and \( p^\dagger \) continue along this direction until they reach a corner of their crowns. At this point, \( p_2 \) and \( p^\dagger \) begin activating their crowns. See Figure \ref{24crowns}.
	 
	Here we compute an upper bound for the makespan of the four-crowns algorithm. By Lemma \ref{crown}, activating \( \mathcal{R}_0 \) and \( \mathcal{R}_1 \) is possible in 
	$$T_{0,1}= r_2 + \|p_2-p_1\|_2 + \|p_1 -t\|_2+ \Big(x + \Big(1+\frac{\phi ^4}{\phi^3 + x} \Big)(1-r_3)\Big)$$
	time units. Also by Lemma \ref{phi}, activating \( \mathcal{R}_2 \) and \( \mathcal{R}_\dagger \) is possible in 
	$$T_{2, \dagger}=r_2 + \|p_2-s\|_2 + \frac{1-r_3}{2} + \Big((\pi-x)+ (1+\phi)(1-r_3)\Big)$$
	time units; we recall that $\phi=\frac{1+\sqrt{5}}{2}$. As the previous algorithms, we set the angle $x$ such that $T_{0, 1}=T_{2, \dagger}$. In other words the wake-up time of four-crowns algorithm is 
	$$\min\limits_{0\leq x\leq\pi} \, \max \{T_{0, 1}, T_{2, \dagger}\}.$$
	
	\vspace{2em}
	
	\textbf{Two-Crowns Algorithm with Respect to $\beta$}
	
	This algorithm is similar to the second strategy in \cite{bonichon:hal-04803161}; the only difference is that both crowns have an angle of \( \pi - \beta \) (instead of \( \pi \)).
	
	Therefore the total time of this algorithm is 
	
	$$r_1 + (\pi - \beta) + \Big(1+\frac{\phi ^4}{\phi^3 + \pi - \beta} \Big)(1-r_1).$$
	Consequently, we consider both the four-crowns and two-crowns algorithms and select the one with the shorter wake-up time.

	\subsection*{\textbf{Algorithms with activating $p_3$ before activating crowns}}
	In some of the previous algorithms, the thickness of the crowns is $1-r_3$
	. Therefore, if $r_3$
	is close to 1, activating the crowns becomes fast. Here, we present some similar algorithms that works better for small values of $r_3$.

	Note that if $r_3$ is small, then $r_1$ and $r_2$ will also be small. As a result, we can apply new algorithms similar to the three-crowns algorithm and four-crowns algorithm.
	
	For more details, as a three-crowns type algorithm, \( p_0 \) can first activate one of \( p_1 \), \( p_2 \), or \( p_3 \). Then, each of the two active robots activates one of the two asleep robots among \( p_1 \), \( p_2 \), and \( p_3 \). Let \( \Omega \) be the angle between these two robots with respect to the origin \( O \). In the next step, each of the four active robots moves by the angle \(\frac{1}{2}|\pi - \Omega| \) to reach two points on the circle \( C(r_3) \). Finally, each of the four robots begins to activate a crown; two of them which are in the same position activate crowns with the angle \( x \), and the other two activate crowns with the angle \( \pi - x \).
	
	With respect to the first activated robot after $p_0$, the previous paragraph gives us 3 algorithms. The fastest one is determined by the positions of $p_1, p_2$ and $p_3$. 
	
	As a three-crowns type algorithm, $p_0$ can first move to one of $p_1, p_2$ or $p_3$; next we can have a strategy similar to the three-crowns algorithm using one of two asleep robots among $p_1, p_2$ and $p_3$. Therefore we have 6 algorithms. Note that in each of these 6 algorithms, the width of the crowns is $1-r_i$, where $\{p_1, p_2, p_3\} - \{p_i\}$ are two activated robots in the three-crowns algorithm.
    \subsection{The Main Algorithm}
For each instance of the Freeze-Tag Problem, we apply all the discussed algorithms and select the one that minimizes the makespan time. In Appendix $\mathbf{B}$, we compute the wake-up time for the main algorithm.

	\section{FTP in $\mathbb{R}^3$}
	\label{section3d}
	
	We provide upper bounds for the wake-up ratio of FTP in $(\mathbb{R}^3, \ell_1)$ and $(\mathbb{R}^3, \ell_2)$ using a similar approach for both cases.

	Given $n$ points inside a unit $\ell_1$-ball ($\ell_2$-ball), we propose an algorithm to wake up the robots. Let $\mathcal{D}$ be the unit $\ell_1$-ball ($\ell_2$-ball) in $\mathbb{R}^3$, and define $\mathcal{D}^{+}$ as the part of $\mathcal{D}$ with positive $z$-coordinates, with $\mathcal{D}^{-} = \mathcal{D} \setminus \mathcal{D}^{+}$. Without loss of generality, assume $\mathcal{D}^{+}$ contains at least as many robots as $\mathcal{D}^{-}$. The algorithm first activates all robots in $\mathcal{D}^{+}$. Then, each robot in $\mathcal{D}^{+}$ activates at most one robot in $\mathcal{D}^{-}$ simultaneously.
	We now present the details for the $\ell_1$-ball and $\ell_2$-ball cases.
	
	\subsection{FTP in $(\mathbb{R}^3,\ell_1)$}
	
	To activate the robots in $\mathcal{D}^{+}$, the active robot $p_0$ moves to the asleep robot with the smallest $z$-coordinate, say $p_1$ (ties are broken arbitrarily). With two active robots at $p_1$, we divide $\mathcal{D}^{+}$ into two regions, assigning one to each robot.
	We will explain in more detail how $\mathcal{D}^{+}$ is divided.
	Two awake robots move toward the asleep robot with the smallest $z$-coordinate in their regions. In each step of the algorithm ties are broken arbitrarily. If no asleep robots are left in a region, the assigned robot stops. When a robot activates another in its region, the region splits, and the two active robots continue the process by handling their parts until all robots are awake.
	
	We explain how the partitioning works. We start with the square $\mathcal{S}$ with corners $(1, 0, 0)$, $(0, 1, 0)$, $(-1, 0, 0)$, and $(0, -1, 0)$. We define a sequence of partitions $\mathcal{Q}_0, \mathcal{Q}_1, \mathcal{Q}_2, \dots$ of the area of $\mathcal{S}$.
	Initially, $\mathcal{Q}_0$ consists of the entire square $\mathcal{S}$.
	At each step, $\mathcal{Q}_i$ is obtained by subdividing each region in $\mathcal{Q}_{i-1}$ into two regions as follows.
	
	For every natural numbers $k$ and $i$ consider the following points in the plane $z = 0$. 
	\vspace{-0.2 cm}
	\begin{eqnarray*}
		e_{ki} = (\frac{i}{2^k}, 1 - \frac{i}{2^k} ),\,\,\,\,\,\,\,\,\,\,\,\,\,\,\,\,\,\,\,\,\,\,\,\,\ e'_{ki} =(\frac{i}{2^k}-1, -\frac{i}{2^k}),\\
		f_{ki} = (-\frac{i}{2^k}, 1 - \frac{i}{2^k}),\,\,\,\,\,\,\,\,\,\,\,\,\,\,\,\,\,\,\,\,\ f'_{ki} = (1-\frac{i}{2^k}, -\frac{i}{2^k}),   
	\end{eqnarray*}
	and 
	\vspace{-0.4cm}
	\begin{eqnarray*}
		E_{k}=\{\overline{e_{ki}e'_{ki}}\,|\,0\leq i\leq 2^k\},\,\,\,\,\,\,\,\,F_{k}=\{\overline{f_{ki}f'_{ki}}\,|\,0\leq i\leq 2^k\}.
	\end{eqnarray*}
	
\begin{figure}
    \begin{subfigure}[b]{0.45\linewidth}
        \resizebox{!}{3.5cm}{

\begin{tikzpicture}[scale=1.6]

\coordinate (E10) at (4,8); 
\coordinate (E11) at (6,6); 
\coordinate (E12) at (8,4); 

\coordinate (Ep10) at (0,4); 
\coordinate (Ep11) at (2,2); 
\coordinate (Ep12) at (4,0); 

\coordinate (F10) at (4,8); 
\coordinate (F11) at (2,6); 
\coordinate (F12) at (0,4); 
\coordinate (Fp10) at (8,4); 
\coordinate (Fp11) at (6,2); 
\coordinate (Fp12) at (4,0); 


    \draw[thick](F10) -- (Fp10);
    \draw[thick](F11) -- (Fp11);
    \draw[thick](F12) -- (Fp12);

    \draw[color = NavyBlue,line width=1.5mm](E10) -- (Ep10);
    \draw[color = NavyBlue,line width=1.5mm](E11) -- (Ep11);
    \draw[color = NavyBlue,line width=1.5mm](E12) -- (Ep12);




    \node at (E10) [yshift = 25, scale=4] {$e_{10}$}; 
    \node at (E11) [xshift = 20,yshift = 25, scale=4] {$e_{11}$}; 
    \node at (E12) [xshift = 20,yshift = 25, scale=4] {$e_{12}$}; 

    \node at (Ep10) [xshift = -8,yshift = -30, scale=4] {$e'_{10}$}; 
    \node at (Ep11) [xshift = -8,yshift = -30, scale=4] {$e'_{11}$}; 
    \node at (Ep12) [xshift = -8,yshift = -30, scale=4] {$e'_{12}$}; 

\end{tikzpicture}
        
        } 
        \caption{$\mathcal{Q}_2$ with 4 squares}
        \label{subfig2}
    \end{subfigure}
    \hspace{0.3cm}
    \begin{subfigure}[b]{0.45\linewidth}
        \resizebox{!}{3.5cm}{

\begin{tikzpicture}[scale=2]
\coordinate (E10) at (4,8); 
\coordinate (E11) at (6,6); 
\coordinate (E12) at (8,4); 

\coordinate (Ep10) at (0,4); 
\coordinate (Ep11) at (2,2); 
\coordinate (Ep12) at (4,0); 

\coordinate (F20) at (4,8); 
\coordinate (F21) at (3,7); 
\coordinate (F22) at (2,6); 
\coordinate (F23) at (1,5); 
\coordinate (F24) at (0,4); 

\coordinate (Fp20) at (8,4); 
\coordinate (Fp21) at (7,3); 
\coordinate (Fp22) at (6,2); 
\coordinate (Fp23) at (5,1); 
\coordinate (Fp24) at (4,0); 


    \draw[thick](E10) -- (Ep10);
    \draw[thick](E11) -- (Ep11);
    \draw[thick](E12) -- (Ep12);

    \draw[color = NavyBlue,line width=1.5mm](F20) -- (Fp20);
    \draw[color = NavyBlue,line width=1.5mm](F21) -- (Fp21);
    \draw[color = NavyBlue,line width=1.5mm](F22) -- (Fp22);
    \draw[color = NavyBlue,line width=1.5mm](F23) -- (Fp23);
    \draw[color = NavyBlue,line width=1.5mm](F24) -- (Fp24);




    \node at (F20) [xshift = -40,yshift = 10, scale=4] {$f_{20}$}; 
    \node at (F21) [xshift = -40,yshift = 10, scale=4] {$f_{21}$}; 
    \node at (F22) [xshift = -40,yshift = 10, scale=4] {$f_{22}$}; 
    \node at (F23) [xshift = -40,yshift = 10, scale=4] {$f_{23}$}; 
    \node at (F24) [xshift = -35,yshift = 10, scale=4] {$f_{24}$}; 

    \node at (Fp20) [xshift = 35,yshift = -12, scale=4] {$f'_{20}$}; 
    \node at (Fp21) [xshift = 35,yshift = -12, scale=4] {$f'_{21}$}; 
    \node at (Fp22) [xshift = 35,yshift = -12, scale=4] {$f'_{22}$}; 
    \node at (Fp23) [xshift = 35,yshift = -12, scale=4] {$f'_{23}$}; 
    \node at (Fp24) [xshift = 35,yshift = -12, scale=4] {$f'_{24}$}; 

\end{tikzpicture}
        
        } 
        \caption{$\mathcal{Q}_3$ with 8 rectangles}
        \label{subfig2}
    \end{subfigure}
    
    \caption{Two partitions of the square $\mathcal{S}$}
    \label{partitionsl1}
    \end{figure}
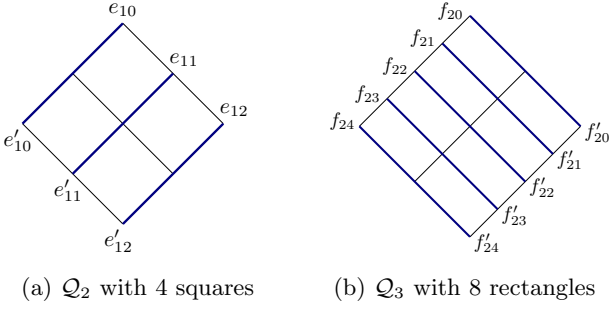

	For every integer $t\geq 1$, the partition $\mathcal{Q}_{2t-1}$ divides $\mathcal{S}$ into $2^{2t-1}$ rectangles using the line segments of $E_{t-1} \cup F_t$. The partition $\mathcal{Q}_{2t}$ then divides $\mathcal{S}$ into $2^{2t}$ squares using the segments of $E_t \cup F_t$. Therefore by definition, $\mathcal{Q}_0$ contains the single square $\mathcal{S}$, and $\mathcal{Q}_1$ consists of two rectangles. See Figure \ref{partitionsl1}.

	In the first step, $p_0$ starts at the origin of $\mathcal{D}$ and is responsible for the entire $\mathcal{D}^+$. It moves toward $p_1$, the asleep robot with the smallest $z$-coordinate in $\mathcal{D}^+$. Using $\mathcal{Q}_1$, we divide $S$ into two rectangles. The planes parallel to the $z$-axis that passes through the segments of $E_1$, split 
	$\mathcal{D}^+$ into two parts; $p_0$ takes one, and $p_1$ takes the other.
	
	In the next step, $p_1$ moves toward $p_2$, the asleep robot with the smallest $z$-coordinate in its assigned part. Once $p_2$ is awake, we partition the corresponding region using $\mathcal{Q}_1$, splitting it into two parts with planes that passes trough segments of $E_1$ and are parallel to the $z$-axis. At this time, $p_1$ is responsible for one part, and $p_2$ for the other.
	This process repeats until all points are awake.
	
	We compute an upper bound for the wake-up time of the algorithm.
	Consider the wake-up tree for the points in $\mathcal{D}^+$. Let $p_0=q_0, q_1, \dots, q_m$ be any arbitrary path from the root to a leaf in the wake-up tree, i.e., in each node of this path there is a robot, either as same as its father or activated by its father. The robot $q_{i+1}$ is activated by one of the $2$ active robots in its parent node.  
	
	We provide an upper bound for the wake-up time along this path.
	Let the coordinates of $q_i$ be $(x_i, y_i, z_i)$. We provide an upper bound for $\sum^{m-1}_{i=0}|x_i-x_{i+1}|+|y_i-y_{i+1}|+|z_i-z_{i+1}|$. In the algorithm, $z_i \geq z_{i-1}$. Since $z_0 = 0$ and the maximum $z$-coordinate in $\mathcal{D}^+$ is at most $1$, it holds that
	\begin{eqnarray}
		\sum^{m-1}_{i=0}|z_i-z_{i+1}|\leq 1. 
		\label{z}
	\end{eqnarray}
	
	Now, we compute an upper bound for $\sum^{m-1}_{i=0}|x_i-x_{i+1}|+|y_i-y_{i+1}|$.

	Let $a_i$ be the projection of $q_i$ onto the plane $z = 0$, i.e., $a_i = (x_i, y_i)$. So, $\|a_i-a_{i+1}\|_1=|x_i-x_{i+1}|+|y_i-y_{i+1}|$.
	In the first step of algorithm, since $q_0$ is at the origin, we have $\|a_0 - a_1\|_1 \leq 1$. For $i\geq 1$,
	when there are two active robots in $q_{i}$, then these two robots should activate the region in $\mathcal{Q}_i$ that contains them. So, the diameter of this region is an upper bound for $\|a_{i}-a_{i+1}\|_1$.
	If $i$ is even then the diameter of each region of $\mathcal{Q}_i$ is $2^{\frac{2-i}{2}}$ and for odd $i$, the diameter of $\mathcal{Q}_i$ is $2^{\frac{3-i}{2}}$. 
	This implies
	\begin{eqnarray*}
		\|a_{i}-a_{i+1}\|_1\leq 
		\begin{cases}
			2^{\frac{2-i}{2}} & \text{for even $i$},\\
			2^{\frac{3-i}{2}} & \text{for odd $i$}.\\
			
		\end{cases} 
	\end{eqnarray*}
	Therefore
    \begin{align}
    \sum_{i=1}^{m-1} \|a_{i} - a_{i+1}\|_1 
    &\leq  \notag \\
    1 + (2 + 2 + 1 + 1 + \tfrac{1}{2} + \tfrac{1}{2} + \dots) 
    &= 9.
    \label{xy}
    \end{align}

	So, by \ref{z} and \ref{xy}, the wake-up time of $\mathcal{D}^{+}$ is at most $9+1=10$. At the final step, since the number of robots in $\mathcal{D}^{+}$ is not fewer than the number of robots in $\mathcal{D}^{-}$, each robot in $\mathcal{D}^{+}$ activates at most one robot in $\mathcal{D}^{-}$.
	Thus, this step requires at most $2$ time units, as the diameter of $\mathcal{D}$ is $2$.
	So, the wake-up ratio of FTP in $(\mathbb{R}^3,\ell_1)$ is at most $12$.
	
	\begin{theorem}\label{norm1}
		$\gamma_{3,1} \leq 12$.
	\end{theorem}

\small
\bibliographystyle{abbrv}


\newpage
\section*{Appendix}
\section*{A: Proof of Lemma \ref{crown}} \label{As}
		Let \( a \) be the active robot at the corner of the inner arc of the crown. Let \( b \) be the closest angular point to \( a \), and let \( \gamma = \angle{aOb}\). First, \( a \) moves along the inner arc at the angle \( \gamma \). See Figure~\ref{x}. This takes \( T_1=(1 - w) \gamma \) time units. Next, \( a \) moves to \( b \), and both active robots follow the direction \( \overrightarrow{Ob} \) until they reach the outer arc of the crown. This takes \(T_2= w \) time units.
		By Lemma \ref{up}, these two robots can activate the rest of the remaining crown \( \mathcal{R}(w, \Theta - \gamma) \) in \( T_3=(\Theta - \gamma) + \left( \frac{\phi^4}{\phi^3 + (\Theta - \gamma)} \right) w \) time units. To prove the lemma it suffices to show that the total time for these steps is at most $\Theta + \left( 1 + \frac{\phi^4}{\phi^3 + \Theta} \right) w.$ We have

		\begin{eqnarray*}
			\frac{\phi ^4}{\phi^3 + \Theta-\gamma} - \frac{\phi ^4}{\phi^3 + \Theta} = 
			\frac{\phi ^4 \gamma}{(\phi^3 + \Theta-\gamma)(\phi^3 + \Theta)} \leq \frac{\phi ^4 \gamma}{\phi^3\phi^3} \leq \gamma.
		\end{eqnarray*}
		So, it holds that 
		\begin{eqnarray*}
			T_1+T_2+T_3=(1-w)\gamma + w +(\Theta-\gamma) + \Big(\frac{\phi ^4}{\phi^3 + \Theta-\gamma}\Big)w  \\ 
			= \Theta+\Big(1-\gamma+\frac{\phi ^4}{\phi^3 + \Theta-\gamma}\Big)w\leq \Theta+\Big(1+\frac{\phi ^4}{\phi^3 + \Theta}\Big)w.\\
		\end{eqnarray*}
	
\section*{B: Computing the Wake-Up Ratio in $(\mathbb{R}^2, \ell_2)$} \label{BBBB}

Due to the mathematical complexity involved in computing the maximum makespan across all our algorithms, we approximated this value using a computational approach.

Consider an input of FTP in \( (\mathbb{R}^2, \ell_2) \). The makespan times of our algorithms are determined by the parameters \( r_1 \), \( r_2 \), \( r_3 \), and the angles \( \mu_{1,2} = \angle p_1 O p_2 \), \( \mu_{1,3} = \angle p_1 O p_3 \), and \( \mu_{2,3} = \angle p_2 O p_3 \). These six values collectively specify the positions of \( p_1 \), \( p_2 \), and \( p_3 \).

It is important to note that, for the two-crowns algorithm, when \( r_2 > 0.87 \), the resulting makespan is given by:

\[
1 + \pi + \left(\frac{\phi^4}{\phi^3 + \pi}\right)(1 - 0.87) \leq 4.27.
\]

Thus, we can assume that \( r_1 \leq r_2 \leq 0.87 \) in this context.

Additionally, note that either \( \mu_{2,3} = \mu_{1,2} + \mu_{1,3} \) or \( \mu_{2,3} = |\mu_{1,2} - \mu_{1,3}| \). We consider both cases for \( \mu_{2,3} \) in the computational program.

The performance of the (two or four)-crowns algorithm is influenced by the value of the angle \( \beta \). Therefore, we consider the maximum makespan over all \( \beta \in [0, \pi] \).

In our computer program \footnote{\href{https://github.com/KB-83/freeze-tag-cccg2025}{Available on GitHub}}, we divide the interval \( [0, 1] \) into 200 equal parts and the angle \( 2\pi \) into 628 equal parts. Specifically, the program computes the makespan time for every \( r_1, r_2, r_3 \in \left\{ 0, \frac{1}{200}, \frac{2}{200}, \dots, 1 \right\} \) with \( r_1 \leq r_2 \leq r_3 \), and every \( \mu_{1,2}, \mu_{1,3} \in \left\{ 0, 0.01, 0.02, \dots, 3.14 \approx \pi \right\} \); see Figure \ref{divide} (a).  For each combination of these parameters, the program computes the minimum makespan time among all our algorithms. 

The final output of the program is the maximum makespan time among all values of \( r_1, r_2, r_3, \beta, \mu_{1,2}, \mu_{1,3} \), and \( \mu_{2,3} \). The output of our computer program is less than 4.2773. In other words,

\[
\max \,\,\min \,\,(\text{makespan time}) \leq 4.2773,
\]

where the maximum is taken over all values of \( r_1, r_2, r_3, \beta, \mu_{1,2}, \mu_{1,3} \), and \( \mu_{2,3} \), and the minimum is taken over all algorithms in Subsection \ref{aaa}.

The positions of the robots are arbitrary within the unit ball, but we consider only a finite set of such positions. Consequently, the approximation of 4.2773 introduces some error. We now present an upper bound for this error.

Our intervals divide the unit ball into 31,500 blocks, but only \( 2 \times 100 \) of these blocks (corresponding to the first and last intervals of the angles) have intersections. Consider an input of FTP, where \( p_1, p_2, \) and \( p_3 \) belong to blocks \( B_1, B_2, \) and \( B_3 \), respectively. By our division, for any block there exists a rectangle with dimensions at most \( \frac{1}{100} \times \frac{1}{200} \). Therefore, for each \( 1 \leq i \leq 3 \), there is a corner \( c_i \) of block \( B_i \) such that \( \| p_i - c_i \|_2 \leq \epsilon \), where

\[
\epsilon = \frac{1}{2} \sqrt{\left( \frac{1}{100} \right)^2 + \left( \frac{1}{200} \right)^2},
\]

See Figure \ref{divide} (b). The main algorithm is as follows: Apply the computer program to the instance \( c_1, c_2, c_3, p_4, p_5, \dots, p_n \); then, add

\[
2\|p_1 - c_1\|_2 + 2\|p_2 - c_2\|_2 + 2\|p_3 - c_3\|_2
\]

to the output (makespan time). This value is added to the makespan because, for each \( 1 \leq i \leq 3 \), in order to activate \( c_i \) with a robot \( q \), the robot \( q \) can move to \( c_i \), then to \( p_i \), and finally return to \( c_i \). These steps add \( \|p_i - c_i\|_2 \) times the unit to the makespan.

Consequently, we have

\[
2\epsilon_1 + 2\epsilon_2 + 2\epsilon_3 \leq \frac{3 \sqrt{5}}{200} \leq 0.0336
\]

as an upper bound for the error of our program. Thus, our total bound is

\[
4.2773 + 0.0336 \approx 4.31.
\]

    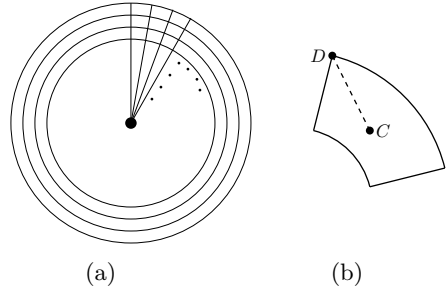
\begin{figure} \label{divide}
    \centering
        \hspace{-0.9cm} 
        \begin{subfigure}[b]{0.3\linewidth} \label{divide}
            \resizebox{!}{3.2cm}{
\begin{tikzpicture}[scale = 1.3]
    \coordinate (O) at (2,2);
    \coordinate (p1) at (2.35,2.4);
    \coordinate (p2) at (2.5,2.6);
    \coordinate (p3) at (2.65,2.8);
    \coordinate (p4) at (2.8,3);
    \coordinate (p5) at (2.95,2.9);
    \coordinate (p6) at (3.1,2.73);
    \coordinate (p7) at (3.15,2.55);

    \draw (2,2) circle (2);
    \draw (2,2) circle (1.8);
    \draw (2,2) circle (1.6);
    \draw (2,2) circle (1.4);

    
    \draw[] (O) -- (2,4);
    \draw[] (O) -- (2.35,3.97);
    \draw[] (O) -- (2.7,3.89);
    \draw[] (O) -- (3,3.75);


    \node at (O) [circle, fill, inner sep=2.5pt] {};
    
    \node at (p1) [circle, fill, inner sep=0.6pt] {};
    \node at (p2) [circle, fill, inner sep=0.6pt] {};
    \node at (p3) [circle, fill, inner sep=0.6pt] {};
    \node at (p4) [circle, fill, inner sep=0.6pt] {};
    \node at (p5) [circle, fill, inner sep=0.6pt] {};
    \node at (p6) [circle, fill, inner sep=0.6pt] {};
    \node at (p7) [circle, fill, inner sep=0.6pt] {};

\end{tikzpicture}

            } 
            \caption{}
            \label{divideB}
        \end{subfigure}
        \hspace{0.5cm} 
        \begin{subfigure}[b]{0.3\linewidth}
            \resizebox{!}{3cm}{
            
\begin{tikzpicture}[scale = 0.8]

\pgfmathsetmacro{\Theta}{14.04}  
\pgfmathsetmacro{\Thetab}{75.96}  

\pgfmathsetmacro{\beta}{11.31}  
\pgfmathsetmacro{\betab}{78.69}  

\pgfmathsetmacro{\r}{4.123}  

    \coordinate (A) at (5,3);
    \coordinate (B) at (4.5,3.2);
    \coordinate (C) at (4,2.4);
    \coordinate (a1) at (1.05,2.45);
    \coordinate (a2) at (1.15,2.5);
    \coordinate (p1) at (4,1);
    \coordinate (p2) at (1,4);
    \coordinate (C) at (2,2);

    \coordinate (p3) at (2,0.5);
    \coordinate (p4) at (0.5,2);
    
    \pgfmathsetmacro{\a}{-2.7}
    \pgfmathsetmacro{\b}{6.75}
    \pgfmathsetmacro{\am}{-1.653}
    \pgfmathsetmacro{\bm}{4.1352}
    \pgfmathsetmacro{\aa}{-4.76}
    \pgfmathsetmacro{\bb}{11.9}
    

    \draw[white] (-1,-1) grid (5,5);

    \draw [thick](p3) arc(\Theta:\Thetab: 0.5 * \r);
    \draw [thick](p1) arc(\Theta:\Thetab:\r);
    
    
    \draw[thick] (p1) -- (p3);
    \draw[thick] (p2) -- (p4);
    \draw[thick,dashed] (p2) -- (C);

    \node at (C) [circle, fill, inner sep=1.7pt] {};  
    \node at (C) [right, scale=1.1] {$C$};  

    \node at (p2) [circle, fill, inner sep=1.7pt] {};  
    \node at (p2) [left, scale=1.1] {$D$}; 

\end{tikzpicture}
            
            } 
           \caption{}
            \label{divideA}
        \end{subfigure}
        \caption{(a) dividing a unit disc into 62800 parts, (b) the distance of each point in a part from its closet corner is at most $|CD|=\epsilon$.}
        \label{divide}
    \end{figure}

	\section*{C: FTP in $(\mathbb{R}^3,\ell_2)$} \label{ha}
	Our approach for $(\mathbb{R}^3,\ell_2)$ is similar to $(\mathbb{R}^3,\ell_1)$.
	Let $\mathcal{D}$ be the unit $\ell_2$-ball in $\mathbb{R}^3$. Define $\mathcal{D}^{+}$ as the part of $\mathcal{D}$ with positive $z$-coordinates, and let $\mathcal{D}^{-} = \mathcal{D} \setminus \mathcal{D}^{+}$.
	Without loss of generality, assume $\mathcal{D}^{+}$ contains at least as many robots as $\mathcal{D}^{-}$.
	
	The initial active robot first activates all robots in $\mathcal{D}^{+}$. Then, each activated robot in $\mathcal{D}^{+}$ activates a corresponding robot in $\mathcal{D}^{-}$.
	
	Now we explain the strategy for activating the robots in $\mathcal{D}^{+}$
	First, $p_0$ moves toward a point $p_1$ with minimum $z$ coordinates in $\mathcal{D}^{+}$ now that there are two active robots, $\mathcal{D}^{+}$ is partitioned into two parts and each of the two robots is responsible for its own part, i.e, it moves toward the asleep robot with minimum $z$ coordinates and activate that robot. We continue this process until all the robots get awake. The way that we partition the sphere is similar as for $(\mathbb{R}^3,\ell_1)$. For integers $k\ge 0$ and $0\le i\le 2^k$
	let 
	\begin{eqnarray*}
		e_{ki} = (-1+\frac{i}{2^{k-1}}, 1, 0),\,\,\,\,\,\,\,\,\,\,\,\,\,\,\,\ e'_{ki} =(-1+\frac{i}{2^{k-1}}, -1, 0),\\
		f_{ki} = (1,-1+\frac{i}{2^{k-1}}, 0),\,\,\,\,\,\,\,\,\,\,\,\,\,\,\,\ f'_{ki} = (-1, -1+\frac{i}{2^{k-1}}, 0),   
	\end{eqnarray*}
	and 
	\vspace{-0.15cm}
	\begin{eqnarray*}
		E_{k}=\{\overline{e_{ki}e'_{ki}}\,|\,0\leq i\leq 2^k\},\,\,\,\,\,\,\,\,F_{k}=\{\overline{f_{ki}f'_{ki}}\,|\,0\leq i\leq 2^k\}.
	\end{eqnarray*}

	Let $\mathcal{S}'$ be a square with the corners $(1, 1, 0), (-1, 1, 0), (-1, -1, 0),$ and $(1, -1, 0)$. We define a sequence of partitioning $\mathcal{Q}_0, \mathcal{Q}_1, \mathcal{Q}_2, \dots$ of the unit $\ell_2$-disk in the plane $z=0$ as follows.
	
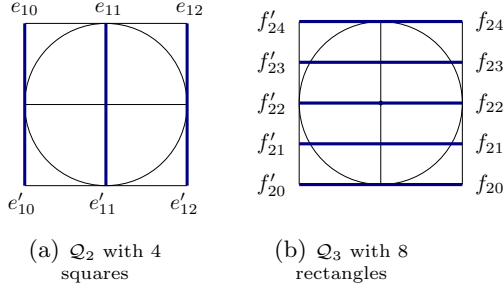
\begin{figure}
    \centering
        \hspace{-0.9cm} 
        \begin{subfigure}[b]{0.3\linewidth}
            \resizebox{!}{3cm}{

\begin{tikzpicture}[scale = 2]
    \coordinate (E10) at (0,8); 
    \coordinate (E11) at (4,8); 
    \coordinate (E12) at (8,8); 
    
    \coordinate (Ep10) at (0,0); 
    \coordinate (Ep11) at (4,0); 
    \coordinate (Ep12) at (8,0); 

\coordinate (F10) at (8,0); 
\coordinate (F11) at (8,4); 
\coordinate (F12) at (8,8); 

\coordinate (Fp10) at (0,0); 
\coordinate (Fp11) at (0,4); 
\coordinate (Fp12) at (0,8); 


    
    \draw (4,4) circle (4);
    
    \node at (4,4) [circle,fill,inner sep = 1.5pt]{};

    \draw[thick](F10) -- (Fp10);
    \draw[thick](F11) -- (Fp11);
    \draw[thick](F12) -- (Fp12);

    \draw[color = NavyBlue,line width=3mm](E10) -- (Ep10);
    \draw[color = NavyBlue,line width=3mm](E11) -- (Ep11);
    \draw[color = NavyBlue,line width=3mm](E12) -- (Ep12);

    \node at (E10) [yshift = 40, scale=6] {$e_{10}$}; 
    \node at (E11) [xshift = 8,yshift = 40, scale=6] {$e_{11}$}; 
    \node at (E12) [xshift = 8,yshift = 40, scale=6] {$e_{12}$}; 

    \node at (Ep10) [xshift = -8,yshift = -50, scale=6] {$e'_{10}$}; 
    \node at (Ep11) [xshift = -8,yshift = -50, scale=6] {$e'_{11}$}; 
    \node at (Ep12) [xshift = -8,yshift = -50, scale=6] {$e'_{12}$}; 

    \end{tikzpicture}

            } 
            \caption{\scriptsize{$\mathcal{Q}_2$ with 4 squares} }
            \label{subfig2}
        \end{subfigure}
        \hspace{0.5cm} 
        \begin{subfigure}[b]{0.3\linewidth}
            \resizebox{!}{3cm}{

\begin{tikzpicture}[scale = 0.65]
\coordinate (E10) at (0,8); 
\coordinate (E11) at (4,8); 
\coordinate (E12) at (8,8); 

\coordinate (Ep10) at (0,0); 
\coordinate (Ep11) at (4,0); 
\coordinate (Ep12) at (8,0); 

\coordinate (F20) at (8,0); 
\coordinate (F21) at (8,2); 
\coordinate (F22) at (8,4); 
\coordinate (F23) at (8,6); 
\coordinate (F24) at (8,8); 

\coordinate (Fp20) at (0,0); 
\coordinate (Fp21) at (0,2); 
\coordinate (Fp22) at (0,4); 
\coordinate (Fp23) at (0,6); 
\coordinate (Fp24) at (0,8); 

    
    \draw (4,4) circle (4);
    
    \node at (4,4) [circle,fill,inner sep = 1.5pt]{};

     \draw[thick](E10) -- (Ep10);
    \draw[thick](E11) -- (Ep11);
    \draw[thick](E12) -- (Ep12);

    \draw[color = NavyBlue,line width=1mm](F20) -- (Fp20);
    \draw[color = NavyBlue,line width=1mm](F21) -- (Fp21);
    \draw[color = NavyBlue,line width=1mm](F22) -- (Fp22);
    \draw[color = NavyBlue,line width=1mm](F23) -- (Fp23);
    \draw[color = NavyBlue,line width=1mm](F24) -- (Fp24);

 \node at (F20) [xshift = 25,yshift = 0, scale=2] {$f_{20}$}; 
    \node at (F21) [xshift = 25,yshift = 0, scale=2] {$f_{21}$}; 
    \node at (F22) [xshift = 25,yshift = 0, scale=2] {$f_{22}$}; 
    \node at (F23) [xshift = 25,yshift = 0, scale=2] {$f_{23}$}; 
    \node at (F24) [xshift = 25,yshift = 0, scale=2] {$f_{24}$}; 

    \node at (Fp20) [xshift = -25,yshift = 0, scale=2] {$f'_{20}$}; 
    \node at (Fp21) [xshift = -25,yshift = 0, scale=2] {$f'_{21}$}; 
    \node at (Fp22) [xshift = -25,yshift = 0, scale=2] {$f'_{22}$}; 
    \node at (Fp23) [xshift = -25,yshift = 0, scale=2] {$f'_{23}$}; 
    \node at (Fp24) [xshift = -25,yshift = 0, scale=2] {$f'_{24}$}; 

    \node at (4,8) [xshift = 0,yshift = 20, scale=1] {}; 

    \node at (4,0) [xshift = 0,yshift = -30, scale=1] {}; 

\end{tikzpicture}

            } 
            \caption{\scriptsize{$\mathcal{Q}_3$ with 8 rectangles}}
            \label{subfig2}
        \end{subfigure}
        \caption{Two partitions of the square $\mathcal{S}'$}
        \label{partitionsl2}
    \end{figure}

	For every integer $t\geq 1$, the partition $\mathcal{Q}_{2t-1}$ divides $\mathcal{S}'$ into $2^{2t-1}$ rectangles using the line segments of $E_{t-1} \cup F_t$. The partition $\mathcal{Q}_{2t}$ then divides $\mathcal{S}'$ into $2^{2t}$ squares using the segments of $E_t \cup F_t$. Therefore by definition, $\mathcal{Q}_0$ contains the single square $\mathcal{S}'$, and $\mathcal{Q}_1$ consists of two rectangles. See Figure~\ref{partitionsl2}.

	We compute an upper bound for the wake-up time of the algorithm.
	Consider the wake-up tree for the points in $\mathcal{D}^+$. Let $p_0=q_0, q_1, \dots, q_m$ be any arbitrary path from the root to a leaf in the wake-up tree, i.e., in each node of this path there are $2$ active robots, one was activated by the other one. The robot $q_{i+1}$ is activated by one of the $2$ active robots in its parent node.  
	We provide an upper bound for the wake-up time along this path, $\sum^{m-1}_{i=0}\|q_{i+1}-q_i\|_2$. Let $q_i=(x_i,y_i,z_i)$, then
		$$\sum^{m-1}_{i=0}\|q_{i+1}-q_i\|_2 $$
        $$=\sum^m_{i=0} \sqrt{(x_i-x_{i+1})^2+(y_i-y_{i+1})^2+(z_i-z_{i+1})^2}$$
        $$\leq \sum^m_{i=0} \sqrt{(x_i-x_{i+1})^2+(y_i-y_{i+1})^2}+|z_i-z_{i+1}|.$$
	According to our algorithm $\sum^{m-1}_{i=0} |z_i-z_{i+1}|\leq 1$. Now, let $a_i=(x_i,y_i)$ be the projection of $q_i$ on the plane $z=0$. So, When $q_i$ moves toward $q_{i+1}$, we only need to compute $\|a_i-a_{i+1}\|_2$, which gives an upper bound for $$\sum^{m-1}_{i=0} \sqrt{(x_i-x_{i+1})^2+(y_i-y_{i+1})^2}.$$
	
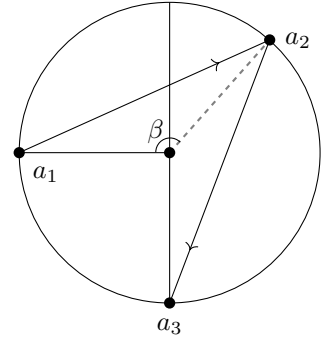
\begin{figure}
    
    \begin{tikzpicture}
    \coordinate (O) at (2,2);
    \coordinate (A) at (0,2);
    \coordinate (B) at (3.33,3.5);
    \coordinate (C) at (2,0);
    \coordinate (a) at (2.14,2.14);

    
    \draw (2,2) circle (2);

     \draw[dash pattern=on 2pt off 2pt, thick, color=
     gray]
    (O) -- (B)
     node[pos=0.5, below right , scale=0.8] {};

    \node at (O) [circle,fill,inner sep = 1.5pt]{};
    \node at (A) [circle,fill,inner sep=1.5pt, label=below right:{ $a_1$}] {};
    \node at (B) [circle,fill,inner sep = 1.5pt,label=right:{ $a_2$}]{};
    \node at (C) [circle,fill,inner sep = 1.5pt,label=below:{ $a_3$}]{};

    \draw[postaction={decorate},
          decoration={markings, mark=at position 0.8 with {\arrow{>}}}] (A) -- (B);
    \draw[postaction={decorate},
          decoration={markings, mark=at position 0.8 with {\arrow{>}}}] (B) -- (C);

    \draw (0,2) -- (O);
    \draw (C) -- (2,4);

    \draw (a) arc (45:180:0.19) node[left, above] {$\beta$};

    \end{tikzpicture}
    \centering
    \caption{The maximum value of $\|a_1-a_2\|_2+\|a_2-a_3\|_2$ is achieved when each of the points  $a_1, a_2,$ and $a_3$  is at a distance of 1 from the origin and $\angle{a_1Oa_3}=\frac{3}{2}\pi$}
    \label{maxalpha}
    \end{figure}

	Without loss of generality assume that $a_1=(x_1,0)$.
	In the first step, since $q_0$ is at the origin, we have $\|a_0 - a_1\|_2 \leq 1$. Next we show an upper bound for $\|a_1-a_2\|_2+\|a_2-a_3\|_2$. Obviously the maximum happens when the robots are on the boundary of $\mathcal{S}'$, assuming that $\angle{a_1a_2a_3}=\alpha$, then $\|a_1-a_2\|_2+\|a_2-a_3\|_2=2\sin \frac{\alpha}{2}+2\sin \frac{3\pi-2\alpha}{4}$. The maximum happens when $\alpha=\frac{3\pi}{4}$. See Figure~\ref{maxalpha}. Thus,

	\begin{eqnarray}
		\|a_1-a_2\|_2+\|a_2-a_3\|_2\leq 4\sin \frac{3\pi}{8}.
		\label{a1a2a3}
	\end{eqnarray}
	For $i\geq 3$,
	when there are two active robots in $q_{i}$, then these two robots should activate the region in $\mathcal{Q}_i$ that contains them. The diameter of this region is an upper bound for $\|a_{i}-a_{i+1}\|_2$.
	If $i$ is even then the diameter of each region of $\mathcal{Q}_i$ is $ 2^{\frac{2-i}{2}}\sqrt{5}$ and for odd $i$, the diameter of $\mathcal{Q}_i$ is $ 2^{\frac{3-i}{2}}\sqrt{2}$. 
	This implies
	\begin{eqnarray*}
		\|a_{i}-a_{i+1}\|_2\leq 
		\begin{cases}
			2^{\frac{2-i}{2}}\sqrt{5} & \text{for even $i$}\\
			2^{\frac{3-i}{2}}\sqrt{2}& \text{for odd $i$}\\
		\end{cases} 
	\end{eqnarray*}
	Therefore
	
		$$\sum^{m-1}_{i=3}\|a_i-a_{i+1}\|_2\leq $$
        \vspace{-0.2cm}
    \begin{eqnarray}
        \sqrt{2}+\frac{\sqrt{5}}{2}+\frac{\sqrt{2}}{2}+\frac{\sqrt{5}}{4}+\cdots=2\sqrt{2}+\sqrt{5}.
		\label{ai}
	\end{eqnarray}
	
	At the final step, since the number of robots in $\mathcal{D}^{+}$ is not fewer than the number of robots in $\mathcal{D}^{-}$, each robot in $\mathcal{D}^{+}$ activates at most one robot in $\mathcal{D}^{-}$.
	Thus, this step requires at most $2$ time units, as the diameter of $\mathcal{D}$ is $2$.
	
	Thus, by \ref{a1a2a3} and \ref{ai}, the wake-up time of $\mathcal{D}$ is at most 
	$$1+1+4\sin \frac{3\pi}{8}+2\sqrt{2}+\sqrt{5}+2 \leq 12.7601.$$ 
	Consequently, the wake-up ratio of FTP in $(\mathbb{R}^3,\ell_2)$ is at most $12.7601$.
	
	\begin{theorem}\label{norm2}
		$\gamma_{3,2} \leq 12.7601$.
	\end{theorem}
    
	Note that our algorithm for computing the upper bound for $\sum^{m-1}_{i=0} \|a_i-a_{i+1}\|_2$ is similar to the algorithm in \cite{YazdiBMK15}, where they give an upper bound of $10.06$ for the wake-up ratio of FTP in $(\mathbb{R}^2,\ell_2)$. But we present an improved analysis, in more details we give a better upper bound on the maximum of $\|a_1-a_2\|_2+\|a_2-a_3\|_2$. This helps to improve the upper bound.

\end{document}